\documentclass{emulateapj}
\usepackage{natbib}
\usepackage{apjfonts}
\usepackage{graphicx}

\newcommand{\lya}{Lyman-$\alpha$ }

\begin{document}

\shorttitle{BAO in the IGM}
\shortauthors{White et al.}

\title{Particle mesh simulations of the Lyman $\alpha$ forest and the
signature of Baryon Acoustic Oscillations in the intergalactic medium}
\author{Martin White${}^{1,2}$,
Adrian Pope${}^{3,4,7}$,
Jordan Carlson${}^1$,
Katrin Heitmann${}^3$,
Salman Habib${}^4$,
Patricia Fasel${}^6$,\\
David Daniel${}^5$,
and Zarija Lukic${}^4$}
\affil{${}^1$ Department of Physics,
University of California Berkeley, CA 94720}
\affil{${}^2$ Department of Astronomy,
University of California Berkeley, CA 94720}
\affil{${}^3$ ISR-1, Los Alamos National Laboratory, Los Alamos, NM 87545}
\affil{${}^4$ T-2, Los Alamos National Laboratory, Los Alamos, NM 87545}
\affil{${}^5$ CCS-1, Los Alamos National Laboratory, Los Alamos, NM 87545}
\affil{${}^6$ CCS-3, Los Alamos National Laboratory, Los Alamos, NM 87545}
\affil{${}^7$ CCS-6, Los Alamos National Laboratory, Los Alamos, NM 87545}
\affil{LA-UR 09-07334}

\date{\today}

\begin{abstract}
  We present a set of ultra-large particle-mesh simulations of the \lya
  forest targeted at understanding the imprint of baryon acoustic oscillations
  (BAO) in the inter-galactic medium.
  We use 9 dark matter only simulations which can, for the first time,
  simultaneously resolve the Jeans scale of the intergalactic gas while
  covering the large volumes required to adequately sample the acoustic
  feature.  Mock absorption spectra are generated using the fluctuating
  Gunn-Peterson approximation which have approximately correct flux
  probability density functions (PDFs) and small-scale power spectra.
  On larger scales there is clear evidence in the redshift space correlation
  function for an acoustic feature, which matches a linear theory template
  with constant bias.
  These spectra, which we make publicly available, can be used to test
  pipelines, plan future experiments and model various physical effects.
  As an illustration we discuss the basic properties of the acoustic
  signal in the forest, the scaling of errors with noise and source
  number density, modified statistics to treat mean flux evolution and
  mis-estimation, and non-gravitational sources such as fluctuations in
  the photo-ionizing background and temperature fluctuations due to
  HeII reionization.
\end{abstract}

\keywords{methods: $N$-body simulations ---
          cosmology: large-scale structure of universe}

\section{Introduction}

Oscillations of the baryon-photon plasma in the early universe, also
known as Baryon Acoustic Oscillations (BAO), imprint a distinct signature
on the clustering of matter \citep{PeeYu70,SunZel70} which provides a
``standard ruler'' by which we can measure the expansion history of the
Universe (see \citealt{EisHu98,MeiWhiPea99} for a detailed description of
the physics in modern cosmologies and \citealt{ESW07} for a comparison of
Fourier and configuration space pictures).  These oscillations have been
traditionally measured in the Cosmic Microwave Background
\citep[CMB; see][for the latest results]{WMAP10}
but with the advent of new, large-volume galaxy redshift surveys BAO have
been detected in galaxy clustering at low $z$ as well
\citep{Eis05,Cole05,Hut06,Bla07,Pad07,Per07,Per09}.

In principle, the BAO technique becomes even more powerful as one moves
to higher redshift, where there is much more volume available to be
surveyed and where the acoustic scale is more deeply in the linear regime.
This latter fact helps in two ways.
First, the power spectrum or correlation function can be computed quite
accurately with only linear perturbation theory once one specifies the
baryon-to-photon ratio and matter-radiation ratio.  These are both measured
accurately from CMB acoustic peaks (see \citealt{WhiCoh02} for a review)
so we have a template with which to fit the data.
Second, the acoustic oscillations are less damped by non-linear evolution
providing more modes which contain measurable signal.

Tracing the enormous volumes required with galaxies requires a heavy
investment in telescope time.
However, in principle any tracer of the mass field will do, including the
neutral hydrogen in the inter-galactic medium \citep{Whi03}
or galaxies \citep{Cha08}.  Tracing neutral hydrogen in galaxies
via its redshifted $21\,$cm emission is a key goal for proposed future radio
telescopes.  However, even with current technology it is relatively
straightforward to obtain a low resolution spectrum of distant quasars and
study the \lya forest of absorption lines which map the neutral hydrogen along
the line-of-sight.
At $z\simeq 2-3$ the gas making up the inter-galactic medium (IGM) is thought
to be in photo-ionization equilibrium, which results in a tight
density-temperature relation for the absorbing material with the neutral
hydrogen density proportional to a power of the baryon density
\citep{HuiGne97,Mei09}.
Since pressure forces are sub-dominant, the neutral hydrogen density closely
traces the total matter density on large scales.
The structure in QSO absorption thus traces, in a calculable way, slight
fluctuations in the matter density of the universe back along the line-of-sight
to the QSO, with most of the \lya forest arising from over-densities of a
few times the mean density.

Motivated by the upcoming Baryon Oscillation Spectroscopic Survey
\citep[BOSS;][a part of SDSS-III]{BOSS}, which will deliver an unprecedented
number of quasar spectra probing the \lya forest at $z\sim 2-3$, and access
to the world's fastest supercomputer, ``Roadrunner'', we have produced a set
of ultra-large particle-mesh simulations of cosmological structure formation
to further investigate BAO-\lya science.
In this paper we describe our simulations, which simultaneously resolve the
Jeans scale and the acoustic scale, and the construction of mock quasar
spectra with properties close to those observed at $z\sim 2-3$
(Sec.~\ref{sec:sim}).
We present several examples to illustrate the use of these spectra in
understanding BAO-\lya science, describing the 2-point correlation function
of the Lyman-$\alpha$ forest flux in Sec.~\ref{sec:twopt}, and the impact
of two non-gravitational signals
(fluctuations in the ionizing background and HeII reionization)
in Sec.~\ref{sec:nongrav}.
We conclude in Sec.~\ref{sec:conclude}.
In the hope that these spectra can be more generally useful, we have made
them publicly available\footnote{{\tt http://mwhite.berkeley.edu}}.

\section{Simulations} \label{sec:sim}

\subsection{Particle-mesh simulations}

The study of the BAO signal in the \lya forest is difficult both analytically
and numerically.  Much of the signal comes from inter-galactic gas at near
mean density, imposing strong requirements on the mass resolution of any
simulation.  The gas contains structure down to the Jeans scale,
$\mathcal{O}(100\,{\rm kpc})$, and resolving this structure while simulating
the extremely large volumes required by BAO demands impressive dynamic range.
As part of the ``Roadrunner Universe'' project \citep{Hab09}
we have run a set of ultra-large particle-mesh simulations, in order
to resolve the Jeans scale in a large cosmological volume.

\begin{figure}
\begin{center}
\resizebox{3.2in}{!}{\includegraphics{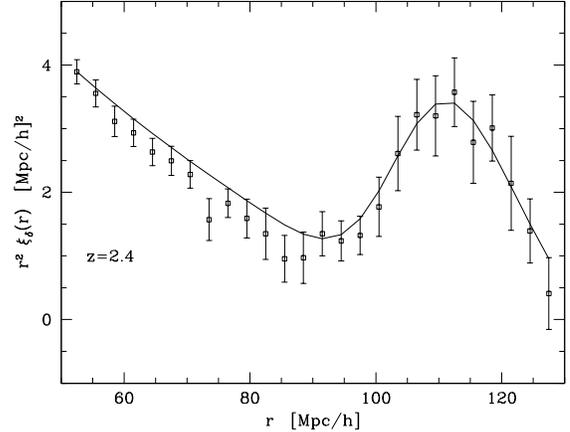}}
\end{center}
\caption{The real-space mass correlation function, $\xi_\delta(r)$, with
bootstrap errors.  The solid line shows the prediction of linear theory,
convolved with a Gaussian of $3\,h^{-1}$Mpc
(see \S\protect\ref{sec:template}).}
\label{fig:ximass}
\end{figure}

In particular we have run $9$ particle-mesh simulations of a flat $\Lambda$CDM
cosmology, with $\Omega_M=0.25$, $\Omega_\Lambda=0.75$, $h=0.72$, $n=0.97$ and
$\sigma_8$ $=0.8$ (i.e.~model 0 of \citealt{Coyote1,Coyote2}).
All of the simulations were started at $z=211$ from an initially regular
Cartesian grid of particles, displaced according to the Zel'dovich
approximation.
Each simulation evolved $4,000^3$ particles in a $750\,h^{-1}$Mpc box
(particle mass $5\times 10^8\,h^{-1}M_\odot$) computing the forces on
a $4,000^3$ grid.
For a description of the particle-mesh code, and code tests, see \citet{Hab09}.
The correlation function of the mass at $z\simeq 2.4$ is shown in
Figure \ref{fig:ximass}, compared to the theoretical prediction.

These simulations are broadly similar to those reported in \citet{SHWL09}.
However those simulations, which were focused on the large-scale BAO signal,
did not properly resolve the small-scale power.  While this should not bias the
shape of the flux correlation function on large scales, it does affect the
distribution of pixels and the large-scale bias, which in turn affects the
signal-to-noise of any measurement.  Further massaging of such low resolution
spectra is thus required before they can be used for pipeline tests, to
investigate signal-to-noise scaling, or as input to Monte-Carlo simulations.
The larger dynamic range in mass and force resolution of the Roadrunner
simulations allows us to maintain high resolution
\citep[comparable to previous authors, e.g.][]{MeiWhi01,McD03,Man03,McD05}
while simulating a relatively large computational volume.
Not even our simulations can resolve $100\,h^{-1}$kpc over the full
$1.5\,h^{-1}$Gpc simulated in \citet{SHWL09}, so we must compromise on either
volume or resolution.
We have chosen to maintain high force and mass resolution by running smaller
boxes than in \citet{SHWL09} and build up total volume by performing several
realizations.

While the total volume can be enlarged by running many individual simulations,
if each has too small a box there can be systematic effects introduced into
the measurement of the acoustic scale.
At the high redshifts of interest for \lya forest work the missing
long-wavelength modes in a finite box simulation do not systematically alter
the evolution of the existing modes --- the fundamental mode of the box is
well in the linear regime.
However the finite support of the $k$-modes implied by periodicity may alter
the correlation function on the scales of interest.
To test for this we generated correlation functions by Fourier transforming
power spectra sampled only on the $k$-grid allowed by periodic boxes of a
range of sizes and compared them to the $L_{\rm box}\to\infty$ limit.  The
largest effect is an almost scale independent suppression of $\xi(r)$ in
the smaller boxes, but this becomes negligible once the box is larger than
$500\,h^{-1}$Mpc.  For our choice of $750\,h^{-1}$Mpc, the correlation
function at $r\sim 10^2\,h^{-1}$Mpc is indistinguishable from that in a
much larger box.
Our choice of box size and resolution is thus adequate for studying the
BAO signature in the \lya forest in some detail.

Recently \citet{NorPasHar09} have reported fully hydrodynamic simulations
covering slightly smaller volumes with slightly worse force and mass resolution
than the simulations reported here.  The advantage of these simulations is
that they attempt to treat the baryonic physics more accurately, but with a
concomitant increase in computational complexity and hence decrease in
dynamic range.

\subsection{Mock spectra}

{}From the phase space data for the particles we produced `skewers' through
the simulation cube of density and line-of-sight velocity at
$z=2.25$, $2.4$, 2.5 and 2.75.  The skewers are distributed at random across
the face of the cube, and run parallel to the sides of the box.


The density and velocity fields were smoothed with a filter of width
$100\,h^{-1}$kpc to approximate the effects of thermal pressure forces
on the gas.  Such a scheme clearly ignores much of the complexity of gas
processes on small scales, but comparison with hydrodynamic simulations
indicates it retains many of the features of interest for the Ly-$\alpha$
forest \citep[see][for a review]{Mei09}.
For an isothermal gas at mean density the (comoving) Jeans filtering scale is
\begin{equation}
  k_J = \sqrt{\frac{3}{2}}\ \frac{H(z)}{(1+z)c_s}
\end{equation}
with $c_s=(5/3\ k_B T/\widehat{m})^{1/2}$ and $\widehat{m}=0.588\,m_H$.
However this is the appropriate smoothing scale only if the gas temperature
scales as $(1+z)$.  While the observational situation is uncertain, it
appears that the gas temperature scales more slowly than this, and the comoving
Jeans scale increases with time.  For this reason \citet{GneHui98} argue for
a smoothing scale set by the Jeans scale at an earlier epoch, specifically
$\simeq 2$ times smaller than the Jeans scale at the time of observation.
Modeling the filtering as a Gaussian and assuming $2\times 10^4\,$K gas
\citep{RicGneShu00,Sch00,McD01,ZalHuiTeg01,The02,Lid09}
gives $\sigma=0.14\,h^{-1}$Mpc at $z=2$ and $\sigma=0.12\,h^{-1}$Mpc at $z=3$.
Given the uncertainties, and the approximate treatment of this physics,
we shall assume $\sigma=100\,h^{-1}$kpc in what follows.

For numerical reasons we used a spline kernel \citep{Deh01} rather than a
Gaussian.  The spline kernel approximates a Gaussian well in the core, but
vanishes identically beyond a range, $R$.  A good match is found when
$R=3.25\,\sigma$, so we adopt $R=0.325\,h^{-1}$Mpc
($\sigma\simeq 100\,h^{-1}$kpc)
which is larger than our mesh scale and mean inter-particle spacing.
We have checked that increasing the resolution does not appreciably change
the spectra when the density field is smoothed on these scales, indicating
that we are numerically converged.

For each of $22,500$ randomly placed lines-of-sight per box the fluctuating
Gunn-Peterson approximation (FGPA; \citealt{CWKH98,GneHui98,Mei09}) was used
to generate skewers of optical depth with $4,000$ pixels each.
We assumed a temperature at mean density of $2\times 10^4$ K \citep{McD01,The02}
and equations of state running from $\gamma=0.5$ to $\gamma=1.5$.
Different choices for the slope, even the inverted equation of state
($\gamma<1$), quantitatively but not qualitatively change our conclusions.
The optical depth included thermal broadening (assumed Gaussian) and
skewers are generated both with and without peculiar velocities for
the gas.

The optical depth was scaled so that the mean transmitted flux
$\bar{F}=\left\langle\exp(-\tau)\right\rangle$ approximately matched that of
the data compiled in \citet{MeiWhi04}:
\begin{equation}
  -\ln\bar{F} = 0.00211\ (1+z)^{3.7}
\label{eqn:Fbarz}
\end{equation}
valid for $1.2<z<4$.
We impose this mean flux condition over the entire volume.
As described later, we also generate skewers in which $\bar{F}$ changes across
the volume to understand the effect of incomplete modeling of the mean flux
evolution.
Another way to set $\bar{F}$ is via the amplitude of the flux power spectrum.
Conveniently, these two methods agreed quite well.
For completeness, we also generate the skewers with dark-matter over-density
only, so we can compare the flux statistics to those of the underlying mass.

We work throughout with relative fluctuations in the flux,
$\delta_F = F(\hat{x})/\bar{F} -1$, so our fundamental data set is
$\delta_F(\vec{x})$ on $22,500$ skewers of $4,000$ pixels each per box.
Our 9 simulations have $V\simeq 3.8\,(h^{-1}{\rm Gpc})^3$ and cover
approximately $1,000$ sq.deg., or 10\% of the coverage planned for BOSS.
On the other hand the line-of-sight areal density ($\sim 200$ per sq.deg.~at
$z=2.5$) is much larger than anticipated from BOSS.
This allows us to study the impact of quasar number density on \lya forest
studies for future missions.
Of course any observational program will likely analyze the data in small
shells of redshift\footnote{Since our simulation boxes are dumped at constant
time they can also be thought of as effectively larger sky area of ``thin''
redshift slices.  However the limited box size means different subregions of
the boxes are not fully independent.} over which the evolution of
e.g.~the mean flux, is small.
A survey such as BOSS should also be able to detect the evolution in $H(z)$
across the redshift range $2<z<3$ by the shift in the acoustic feature in
velocity space.

\subsection{Properties of the simulated spectra}

At the fiducial redshift of our box, $z\sim 2.5$, the Ly-$\alpha$ feature
is redshifted to approximately $4,000\,$\AA, and the box has a velocity
``length'' of $73,000\,$km/s.
Each of our mock spectra thus encompasses the full Ly-$\alpha$ to Ly-$\beta$
region for QSOs at $z\sim 2.5$ (Table \ref{tab:conversion}).
For $z\simeq 2-3$ a comoving $h^{-1}$Mpc is approximately equal to
1\AA\ (Table \ref{tab:conversion}),
so with $4,000$ pixels per spectrum, each simulation pixel is $0.2\,h^{-1}$Mpc
wide.  For comparison each SDSS-III pixel is about $75\,$km/s (or just under
$1\,h^{-1}$Mpc), so our spectra are comparably well resolved.

\begin{table}
\begin{center}
\begin{tabular}{ccccccc}
    $z$ & $\lambda_\alpha$ & $\chi_\alpha$ & $\chi_\beta-\chi_\alpha$
    & $H(z)$ & $d\lambda/d\chi$ & $dv/d\chi$ \\ \hline
   1.75 &  3343.1 & 3506 &   590 &    244  &    0.99 & 89 \\
   2.00 &  3647.0 & 3796 &   575 &    274  &    1.11 & 91 \\
   2.25 &  3950.9 & 4055 &   561 &    305  &    1.24 & 94 \\
   2.50 &  4254.8 & 4288 &   546 &    339  &    1.37 & 97 \\
   2.75 &  4558.8 & 4499 &   531 &    373  &    1.51 & 100 \\
   3.00 &  4862.7 & 4690 &   518 &    409  &    1.66 & 102
\end{tabular}
\end{center}
\caption{Some useful conversion factors for our cosmology.
For each redshift, $z$, we list the observed wavelength [$\lambda_\alpha$]
of Ly-$\alpha$ emitted at that $z$, the comoving distance to that redshift
[$\chi_\alpha$], the extra comoving distance to Ly-$\beta$ if Ly-$\alpha$
is at that $z$ [$\chi_\beta-\chi_\alpha$], the Hubble parameter [$H(z)$]
and conversions between distance and wavelength [$d\lambda/d\chi$] and
velocity [$dv/d\chi$].  Distances are measured in $h^{-1}$Mpc, velocities in
km/s and wavelengths in \AA.}
\label{tab:conversion}
\end{table}

\begin{figure}
\begin{center}
\resizebox{3.2in}{!}{\includegraphics{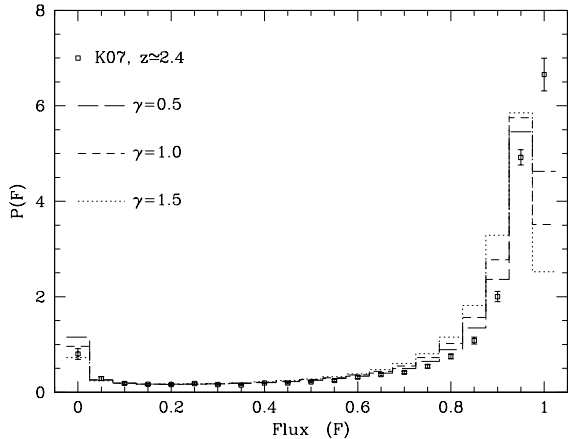}}
\end{center}
\caption{The flux PDF in simulations (lines) and as measured in high resolution
spectra by \protect\citet[][squares]{Kim07}.  Overall the simulations do a
fair job of describing the flux PDF, containing approximately the right
distribution of low and high absorption regions.}
\label{fig:Fpdf}
\end{figure}

We normalize the optical depths so that the mean transmitted flux
matches observations.   An additional constraint then comes from
comparing the flux variance of the simulations and observations.
For $\gamma=0.5-1.5$ we find  $\sigma_F^2=0.1$, which is very comparable
to the measurements reported in \citet{McD00}.
Figure \ref{fig:Fpdf} shows the flux PDF in the simulations compared to that
measured from high resolution spectra at $z\simeq 2.4$ in \citet{Kim07}.
We see that the distribution of high and low absorption regions is
approximately correct, indicating that the spectra could be useful for
testing analysis pipelines and investigating the impact of systematic
errors.  While the discrepancies are larger than the quoted observational
error bars, we note that our modeling could be improved and the flux PDF
is notoriously difficult to measure observationally
(especially in regions of low absorption).
We have not attempted a more detailed comparison since the level of agreement
will be sufficient for our purposes.

\begin{figure}
\begin{center}
\resizebox{3.0in}{!}{\includegraphics{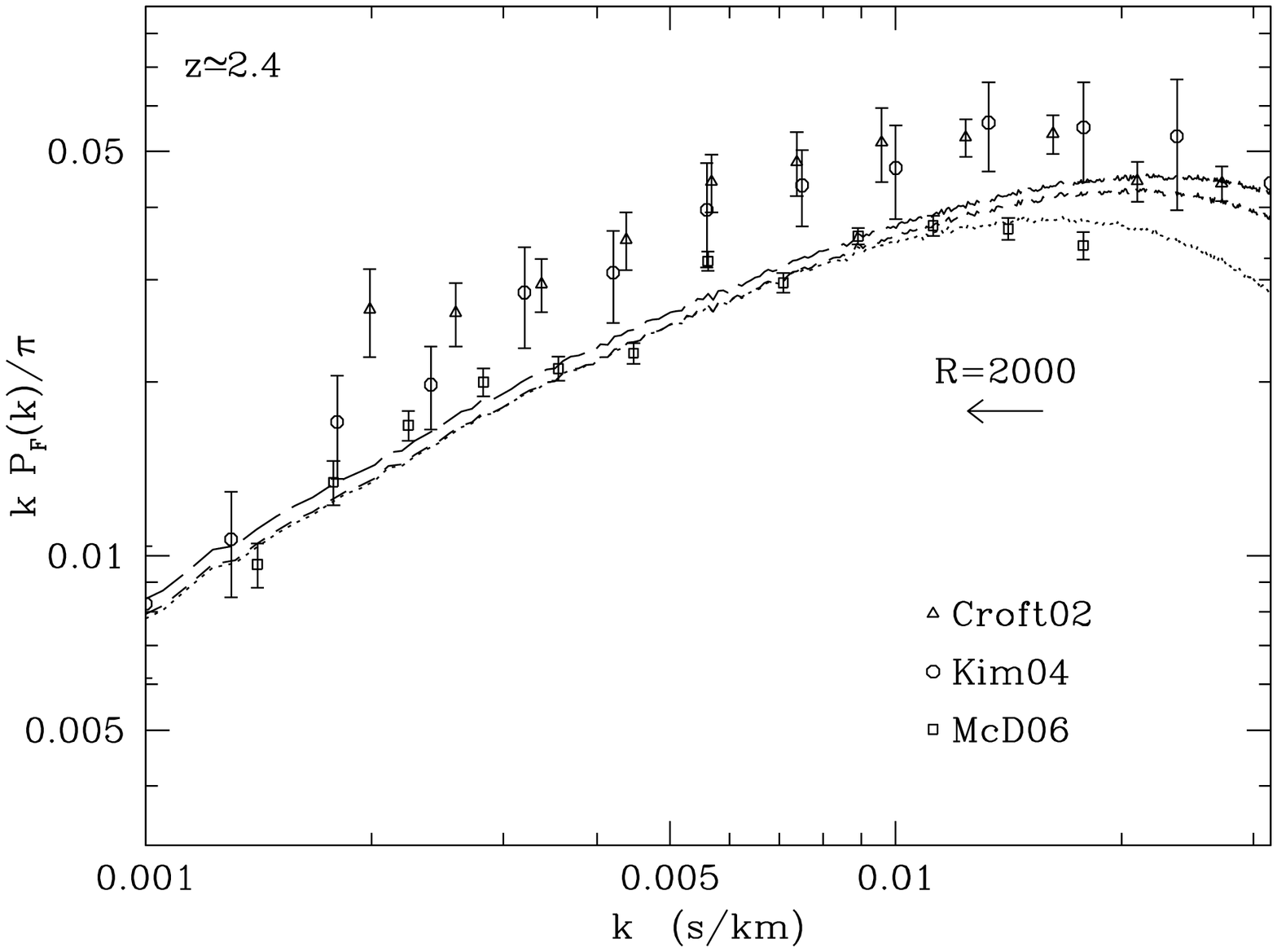}}
\end{center}
\caption{The (dimensionless) line-of-sight power spectrum of the flux
measured from the simulations at $z\simeq 2.4$ and from
\protect\citet[][triangles]{Croft02}, \protect\citet[][circles]{Kim04} and
\protect\citet[][squares]{McD06}.
As in Figure \protect\ref{fig:Fpdf}, line types indicate the assumed
equation-of-state: $\gamma=0.5$ (long-dashed), $\gamma=1.0$ (short-dashed) and
$\gamma=1.5$ (dotted).  The arrow indicates roughly the range of scales
accessible from $R=2,000$ spectra such as will be produced by BOSS.}
\label{fig:SmallPk}
\end{figure}

The line-of-sight power spectrum of the flux at $z\simeq 2.4$, compared to
the measurements of \citet{Croft02}, \citet{Kim04} and \citet{McD06},
is shown in Figure \ref{fig:SmallPk}.
There is some tension between the flux PDF and the power spectrum in the
preferred value of $\bar{F}$: the flux PDF is better fit if we slightly
raise $\bar{F}$, increasing the number of lines of sight with little or
no absorption and decreasing the number with little or no transmission.
The flux power spectrum is better fit if we lower $\bar{F}$, raising the
amplitude of $P_F(k)$.  For $\bar{F}\simeq 0.8$ the overall the level of
agreement is good in both statistics, indicating that the relative amounts
of power on different scales are approximately as observed in the real
Universe.  We have chosen to match to the later, SDSS-based measurements of
\citet{McD06} rather than the slightly higher results of \citet{Croft02,Kim04}.
Our spectra thus contain approximately the right distribution of fluxes and
approximately the right amount of small-scale structure, which acts as a
source of ``noise'' in the measurement of the acoustic scale.

In what follows we will work with $\delta_F$, ignoring real-world issues
such as continuum fitting or subtraction, damped \lya systems and metal
lines.  We expect the damping wings and metal lines to be uncorrelated
with the signal of interest, and so not produce a feature at the acoustic
scale.
Similarly, we will be looking at cross-correlation statistics between
different lines-of-sight.  Since several million QSO pairs contribute
to the flux 2-point function, $\xi_F(r_{BAO})$, the fluctuations will
rapidly average to zero \citep{Vie02}.

\section{Two point statistics} \label{sec:twopt}

\subsection{The correlation function}

We compute the density-density or flux-flux (cross-) correlation function as
a direct sum over pixel pairs, omitting pixel pairs in the same line-of-sight.
We will focus on isotropically averaged statistics -- due to our limited
volume we do not have a statistically significant detection of the quadrupole
[$\xi_2$] in the acoustic peak region\footnote{We do see evidence at smaller
$r$ for a lower quadrupole-to-monopole ratio than in the case of the mass.
This is expected due to the mapping from density to flux.}.
We use $3\,h^{-1}$Mpc bins, such that the effect of binning changes $\xi$ by
less than $1\%$ in the region of the acoustic peak.  The binned correlation
function for any model can be as easily computed as the unbinned correlation
function, e.g.
\begin{equation}
  \xi_F(r_i;\Delta r) = \int \frac{d^3k}{(2\pi)^3}\ P_F(k) W_i(k)
\end{equation}
where $P_F$ is the flux power spectrum and
\begin{equation}
  W_i(k) = 3\ \frac{\sin(x_{+})-\sin(x_{-})-x_{+}\cos(x_{+})+x_{-}\cos(x_{-})}
           {x_{+}^3-x_{-}^3}
\label{eqn:wk}
\end{equation}
with $x_\pm=k(r\pm \Delta r/2)$.
As $\Delta r\to 0$ we have $W_i\to j_0(kr)\equiv \sin(kr)/(kr)$.
Larger $\Delta r$ leads to increased damping as $k\to\infty$ and a very
slight increase in the degree of correlation between adjacent bins.
For $\Delta r\to 0$, infinitesimal pixels and white noise, the error on a
given $\xi_F(r)$ is infinite.  The noise contribution to a binned
statistic is however finite.
If our $r$ bins are larger than our pixels (as in our case) then it is
the bin width which tames the noise -- any results on e.g.~S/N
are thus dependent on the bin size.


\subsection{Template} \label{sec:template}

\begin{figure}
\begin{center}
\resizebox{3.2in}{!}{\includegraphics{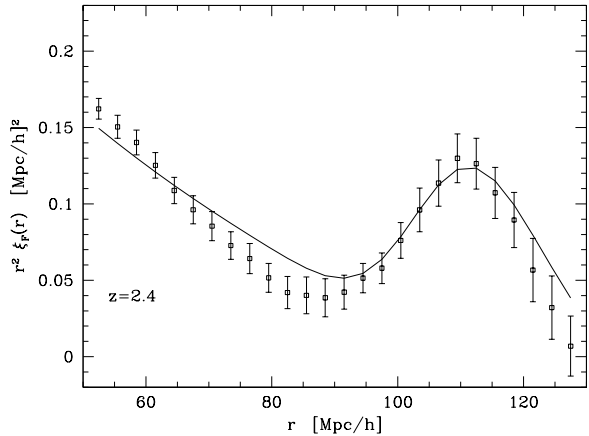}}
\end{center}
\caption{The redshift space flux correlation function, $\xi_F$, with
bootstrap errors.  The solid line shows linear theory, convolved with a
Gaussian and multiplied by a constant bias of $b^2=0.2^2$
(see Figure \protect\ref{fig:bias}).}
\label{fig:xiF}
\end{figure}

In general the flux correlation function (Figure \ref{fig:xiF}) can be
related to the non-linear, redshift-space, mass correlation function,
$\xi_m(r)$ as
\begin{equation}
  \xi_F(r) = B(r) \xi_m(r) + A(r)
\end{equation}
for some smooth functions $B(r)$ and $A(r)$.

A first approximation to $\xi_m$ is to take a multiple of the real-space mass
correlation function predicted by linear theory.
The dominant effect of non-linear clustering is to broaden the acoustic
peak, with an amplitude that can be estimated from the rms Zel'dovich
displacement \citep{Bha96,ESW07,CroSco08,Mat08}.
At $z=2.5$ this is about $3\,h^{-1}$Mpc in our cosmology, to be compared to
the much larger intrinsic width of the acoustic feature (set by the diffusion,
or Silk, damping scale: $12\,h^{-1}$Mpc).  Figure \ref{fig:ximass} shows the
mass correlation function measured in the simulations (with errors determined
{}from bootstrap as described below) compared to this smoothed
linear theory.  The agreement is very good over a broad range of scales
(recall the errors are highly correlated).

On the large scales of interest here redshift space distortions simply
multiply the real-space, mass correlation function by $1+(2/3)f+(1/5)f^2$
-- where $f\equiv d\ln\delta/d\ln a\simeq \Omega_m^{0.6}\simeq 1$ is the
growth factor \citep{Kai87,Ham92} -- and apply smoothing of the same order
as above.
Adding these in quadrature we see non-linear evolution and redshift space
distortions will only change the peak width by $\sim 5\%$.
A reasonable approximation to $\xi_m(r)$, therefore, is the linear theory,
real-space, mass correlation function multiplied by a constant and convolved
with a Gaussian of $4\,h^{-1}$Mpc.

\begin{figure}
\begin{center}
\resizebox{3.2in}{!}{\includegraphics{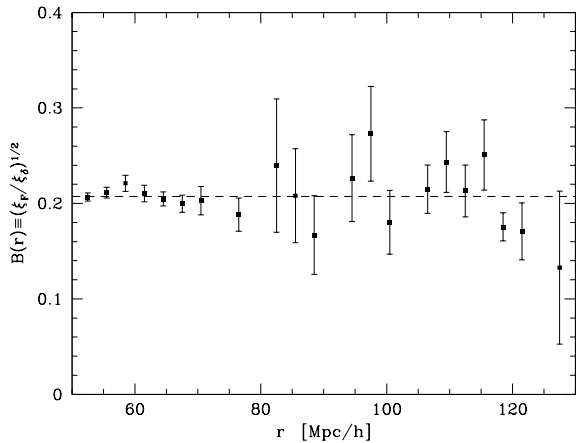}}
\end{center}
\caption{The flux bias, $B^2(r)\equiv \xi_F/\xi_\delta$, as a function of
scale.  While the results are noisy, the bias is consistent with being
scale-independent as noted earlier by \protect\citet{SHWL09}.}
\label{fig:bias}
\end{figure}

\citet{SHWL09} found $B(r)$ was approximately scale-independent, suggesting
a good template for the flux correlation function is the linear theory,
real-space, mass correlation function multiplied by an effective bias --
shown as the solid line in Figure \ref{fig:xiF}.
Our simulations are consistent with this finding (Figure \ref{fig:bias}),
but since we have less volume our constraints are not as strong.

\subsection{Covariance} \label{sec:cov}

To assign error bars to the points we can use either the box-to-box scatter or
a bootstrap scheme \citep{Efron82}.
We tried two different methods for determining the bootstrap errors.  The first
was simply to draw boxes randomly (with replacement) and compute the average
correlation function from the boxes.  In the second we randomly assigned each
skewer to one of $10$ bunches within each simulation and used bootstrap
resampling on the full collection of bunches in all of the boxes.
Within any individual box the bootstrap errors will drastically underestimate
the covariance on large scales, however in the limit that we have a large
number of independent boxes and the skewer bunches are uncorrelated it will
provide a good estimate of the error from the finite survey volume (which we
expect to dominate).  We find that the box-to-box variance and first bootstrap
scheme provided consistent and more stable indicators of the large-scale errors
than the second bootstrap.  This suggests that with observational data one
wants to bootstrap on disjoint regions of the sky rather than on individual
quasar sight-lines.  It may also be helpful to apply a high-pass filter to
make the individual regions more independent
\citep[see e.g.][for discussion]{PWNP09}, though we do not need to do this
in our simulations.

\begin{figure}
\begin{center}
\resizebox{3.2in}{!}{\includegraphics{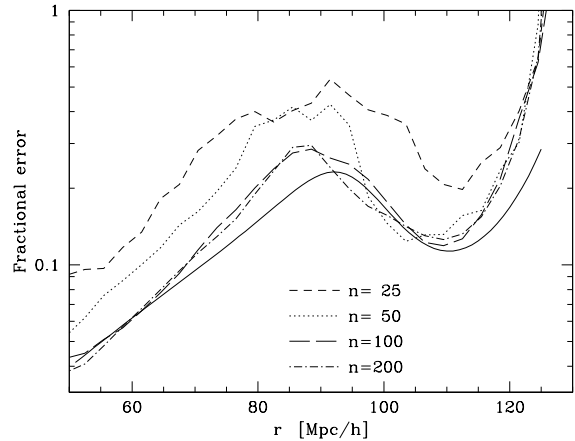}}
\end{center}
\caption{The fractional error on $\xi_F$ as a function of scale for different
choices of sight-line areal density (per sq.deg.).
Since the errors can be noisy, we smoothed them over a bin of width
$10\,h^{-1}$Mpc.  Note the fractional error is larger when $\xi_F$ is small,
near $r\simeq 90\,h^{-1}$Mpc, and smaller near the peak in $\xi_F$ at
$r\simeq 110\,h^{-1}$Mpc.
The solid line shows the expectation for Gaussian fluctuations.}
\label{fig:error}
\end{figure}

Our 9 boxes cover approximately $1,000$ sq.deg., or roughly 10\% of the
area planned for BOSS.  In this regime the errors scale as area${}^{-1/2}$,
so if all BOSS quasars probed the \lya forest at a fixed redshift, and if
noise were not an issue, it should achieve errors $\simeq 3\times$ smaller
than those plotted in Figure \ref{fig:error} at similar number densities.

Using this scheme we can look at the dependence of the covariance on the number
of lines-of-sight per square degree and on the noise in the individual
spectra.  Here we vary each of these degrees of freedom independently, allowing
us to disentangle the contributions from finite volume, finite sampling and
noise.  Observationally the only way to achieve a high number density of
sources is to go further down the luminosity function, thus at fixed exposure
time the signal-to-noise will generally be lower at higher number density.
Furthermore the signal-to-noise will usually vary across the spectrum.  A
detailed study of these issues, or an optimization for a given system, is
beyond the scope of this paper.

Figure \ref{fig:error} shows the dependence of the error on $\xi_F$ on the
density of skewers.  Since the error on the error is quite large from our
small number of simulations we have smoothed the error with a sliding window
of 3 bins in making this figure.  Figure \ref{fig:error} shows that
increasing the areal density of sight-lines from 25 to 50 per sq.deg.~results
in better determination of $\xi_F$ in the acoustic peak region.  However the
gains saturate between 50 and 100 quasars per sq.deg.~and going to 200 quasars
per sq.deg.~results in no improvement as the error is dominated by the finite
volume surveyed.

These trends can be readily understood by considering the case of a Gaussian
random field which is sampled along a large but finite number of `skewers'.
The observed power spectrum is related to the underlying 3D power spectrum
through a window function.
For a single $\mathbf{k}$-mode the variance is unaffected by this step, but
when considering angularly averaged quantities [e.g.~$P(|k|)$ or $\xi(r)$]
the error is increased \citep{MeiWhi99,HamRimSco06}.
In the limit that the skewers are placed randomly one can show that
(see Appendix)
\begin{equation}
  {\rm Var}\left[P_F\right] = \sum_{j=0}^3 \frac{v_j}{\bar{n}^j}
\label{eqn:Vjdef}
\end{equation}
where $\bar{n}$ is the number density of sightlines and the $v_j$ involve
the power spectrum and its integral in a $\mathbf{k}$ shell,
$k-\Delta k/2<|\mathbf{k}|<k+\Delta k/2$.  The lowest order term,
$v_0=2P_F^2/N_k$, is the familiar result with $N_k$ the number of
modes in the shell \citep{FKP}.
The ratio $v_1/v_0$ controls the number density at which we achieve essentially
sample variance limited constraints, with the relevant figure of merit
depending on the ratio of the achieved line-of-sight (i.e.~2D) quasar number
density to a ``critical'' value (near $0.01\,h^2\,{\rm Mpc}^{-2}$)
as described in the Appendix.  For observations at $z\sim 2$ this critical
number density is between $50-100$ quasars per square degree (depending on
the redshift range considered), which is where we see diminishing returns in
Figure \ref{fig:error}.

While quasar counts at such faint limits are quite uncertain, the critical
number density corresponds to quasars brighter than a $B$-band magnitude of
about 22.
Since it is unlikely that all such quasars in the range $2<z<3$ could be
successfully targeted, and since not all of the forest region will be useful
for cosmology in all quasars, the relevant sample limits should be fainter.
At these magnitudes going one magnitude fainter approximately doubles the
number of quasars per square degree (in a fixed, depth $\sim 500\,h^{-1}$Mpc,
redshift slice) and going two magnitudes fainter increases the density by
a factor $\sim 3.5$.

In the limit of infinite sampling density the error approaches the sample
variance limit, which for $\xi(r)$ can be written as
\begin{equation}
  {\rm Cov}\left[\xi_i,\xi_j\right] = 
  \frac{2}{V}\int \frac{d^3k}{(2\pi)^3}\ P_F^2(k)\, W_i(k)W_j(k)
\end{equation}
where $W_i$ indicates the window function for bin $i$ (Eq.~\ref{eqn:wk}) and
we have assumed the monopole dominates the higher multipoles.
The relevant volume is $V=3.8\,(h^{-1}{\rm Gpc})^3$ and the plotted error
in Figure \ref{fig:error} is the square root of the diagonal entries of
this matrix.
For a large number of skewers the simulation-based errors are in reasonable
agreement with the Gaussian prediction in the region of the acoustic peak,
however the results are still somewhat noisy.  The results become increasingly
noisy for the off-diagonal elements of the covariance matrix, so we have not
attempted a quantitative comparison.
The Gaussian prediction for our $3\,h^{-1}$Mpc bins is that $\xi$ at the peak
($r\simeq 110\,h^{-1}$Mpc) is correlated with $r=50$, 90 and $100\,h^{-1}$Mpc
at the 30, 80 and 90 per cent level.

\begin{figure}
\begin{center}
\resizebox{3.2in}{!}{\includegraphics{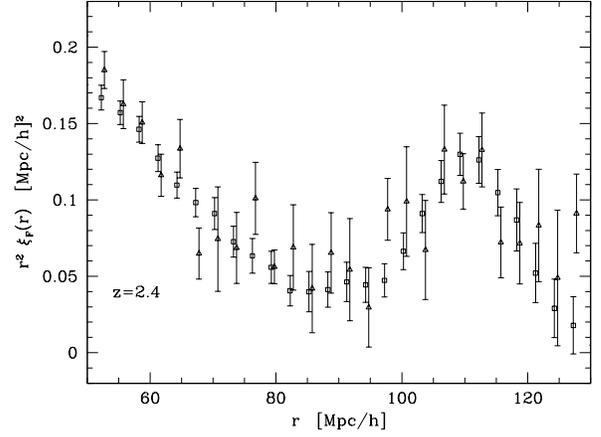}}
\end{center}
\caption{The redshift space, flux correlation function for the spectra with
no noise (squares) and including noise with $S/N=1$ per $h^{-1}$Mpc (which
is close to $1$ per \AA; triangles).  The errors are estimated by bootstrapping
the 9 boxes.}
\label{fig:noise}
\end{figure}

Figure \ref{fig:noise} shows the effects of adding noise to the spectra.
We use $100$ quasars per sq.deg., since at that areal density the volume
is well sampled, and add uncorrelated Gaussian noise to each pixel such
that the $S/N$ is unity per $h^{-1}$Mpc, or very close to unity per \AA.
It is clear from Figure \ref{fig:noise} that noise at this level is a
non-trivial contributor to the total error budget at this number density
and volume.  Recall, as discussed above, that in the absence of binning
or finite pixels the white-noise error on $\xi$ at any $r$ is infinite.
Since our $r$-bins are larger than our pixels it is this size which tames
this infinity, so the precise results depend on the chosen binning.

\subsection{Broad band power \& compensated statistics}
\label{sec:deltaXi}

One of the advantages of measuring the correlation function, rather than the
power spectrum, is that it is straightforward to compute for even complicated
geometries \citep{SHWL09}.  However when estimating the correlation function
one needs to pay attention to the mean flux in the survey (equivalent to the
mean galaxy density, or ``integral constraint'' in the case of galaxies).
This is not an issue for estimates of the power spectrum, which neatly
immunizes high $k$ power from slowly varying signals such as a mean-density
or mean-flux error.  A (small) misestimate of $\bar{F}$ in the power spectrum
leads to a (small) misestimate of the amplitude of $P_F$.
In the correlation function however an error in the mean flux generates both
an additive and multiplicative error,
\begin{equation}
  \xi_F\to \frac{\bar{F}^2_{\rm true}}{\bar{F}^2_{\rm est}}\xi_F +
  \left(1-\frac{\bar{F}_{\rm true}}{\bar{F}_{\rm est}}\right)^2
\end{equation}
and the additive term can swamp the desired correlation function at large
scales.
If the mean flux evolves with redshift (or distance) the situation is slightly
more complicated.  However the limit where the mean flux for any pair of
points is the same (e.g.~the separation of the two points is transverse to the
line of sight or the separation vector is much smaller than the characteristic
scale over which $F$ varies) is a simple generalization of the above.
If $\xi_F$ is $\chi-$ or $z$-independent (as it is in our boxes) and the
measured mean flux is $\bar{F}$, we have
\begin{equation}
 \xi_{\rm obs}(r) = \frac{\langle F^2\rangle}{\bar{F}^2}\ \xi_F
  + \left(1+\frac{\langle F^2\rangle}{\bar{F}^2}
          -2\frac{\langle F\rangle}{\bar{F}} \right)
\end{equation}
where $\langle\cdots\rangle$ indicates an average over distance or redshift.
The general case is more complex, involving additional correlators of the
mean flux and $\delta_F$.

To handle this one can either fit out a smoothly varying piece of $\xi_F$
\citep[as advocated in e.g.][to handle galaxies]{SeoEis05,SeoEis07,Whi05}
or compute a statistic which automatically reduces the sensitivity to
slowly varying modes
\citep[e.g.][]{HSWSW07,PadWhiEis07,OmEll}.

\begin{figure}
\begin{center}
\resizebox{3.2in}{!}{\includegraphics{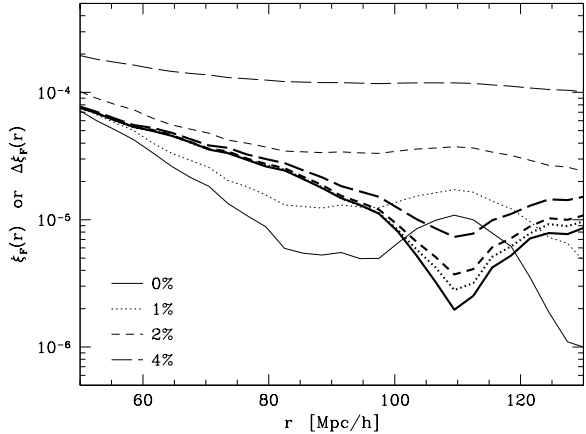}}
\end{center}
\caption{The flux correlation function, $\xi_F$ (thin lines), and
compensated correlation function, $\Delta\xi_F$ (thick lines), for situations
where the mean flux varies linearly across the box with an end-to-end
amplitude of $0$, $0.01$, $0.02$ and $0.04$ about a fixed mean flux at the
box mid-plane.}
\label{fig:evolve}
\end{figure}

Figure \ref{fig:evolve} shows the ``extra'' correlation that is introduced
by a mean flux which varies linearly (by $0$, $1$, $2$ and $4\%$) with
distance across the box and that much of this is taken out by
\begin{equation}
  \Delta\xi_F(r) \equiv \left\langle \xi_F(<r) \right\rangle - \xi_F(r)
\end{equation}
where the first term indicates the average of the correlation function within
$r$ \citep{HSWSW07}.
Note that in $\Delta\xi_F$ the acoustic ``peak'' becomes an acoustic ``dip''.
We found that the change in $\Delta\xi_F$ with increasing mean flux
variation is not well fit by a constant multiplicative or constant
additive factor.
Interpreted as a multiplicative change, the ratio is $1$ below
$\sim 50\,h^{-1}$Mpc, grows smoothly to a factor of 1.4(4) at the peak,
and then drops again beyond the peak for mean flux evolution of 1(4)\%
across the box.
The larger the mean flux evolution, the broader the effect in $r$ and for
all cases the ratio peaks at the acoustic dip ($r\simeq 110\,h^{-1}$Mpc)
and thus alters the dip location very little.
If unaccounted for such evolution would lead to poor fits between the
theoretical template and the observations, but the acoustic signal survives
in the measurement.

Assuming Eq.~(\ref{eqn:Fbarz}), to limit the flux evolution to less than 1\%
we would need to analyze the spectra in shells of redshift $\Delta z\le 0.05$
and for 4\% the shell would have to be $\Delta z\le 0.2$.
Alternatively we could imagine including a model for the mean flux evolution
in our theoretical prediction or, since we know the mean flux evolution is in
the line-of-sight direction it may be possible to isolate this piece by
appealing to a model for the angular dependence of $\xi_F(\vec{r})$.
Since such a model should also handle redshift space distortions we defer
this to a future publication.

\section{Non-gravitational contributions} \label{sec:nongrav}

One possible cause for concern, in addition to systematic errors in the
measurements themselves, is non-gravitational contributions to the flux
correlations.  These could arise from hydrodynamic forces, radiative
transfer effects, reionization heating or other departures from the simple
FGPA assumed thus far \citep{Mei09}.
Many of these effects are expected to contribute mostly on small scales,
with no power preferentially on the acoustic scale, and should not bias
a BAO measurement (though it may reduce the signal-to-noise of the measurement).
Here we investigate two such non-gravitational contributions:
spatial fluctuations in the ionizing background radiation and
spatial fluctuations in the temperature-density relation of the gas
as could arise e.g.~from (inhomogeneous) HeII reionization.
As mentioned previously, fluctuations in the quasar continuum should be less
of a concern when we average the cross-spectra from many different quasar
sightlines than when we look at the auto-spectra of individual sightlines
\citep{Vie02,Whi03}.

\subsection{Fluctuations in the ionizing radiation}

One possible contributor of large-scale power is fluctuations in the UV
background field or the photo-ionization rate ($\Gamma$).
Since the attenuation length of the IGM at $z\sim 2-3$ is
$\mathcal{O}(100\,{\rm Mpc})$, and the background is thought to be
dominated by rare sources (QSOs), $\Gamma$ may have spatial structure on
large scales \citep{Zuo92,FarShu93,GneHam02,MeiWhi04,Croft04,McD05}.
Assuming the IGM is in photo-ionization equilibrium the optical depth,
$\tau\propto\Gamma^{-1}$.
\citet{SHWL09} investigated the effects of a fluctuating photo-ionization
rate in a highly simplified model.
We follow \citet{SHWL09} in assuming that the ionizing background field is
dominated by light from quasars, but extend the analysis to consider
diagnostics of these additional fluctuation sources.

We place our quasars at random within the volume with luminosities drawn from
a luminosity function
\begin{equation}
  \Phi \propto \frac{1}{(L/L_\star)^{-\alpha}+(L/L_\star)^{-\beta}}
\end{equation}
with $\alpha=-3.31$ and $\beta=-1.09$ \citep{Cro04}.
Each QSO is assumed to emit isotropically with constant luminosity
$L$, so the contribution to the photo-ionization rate from the $i$th QSO at
distance $r_i$ can be taken to be
\begin{equation}
  \Gamma_i \propto L_i\ \frac{e^{-r_i/r_0}}{4\pi r_i^2} \quad ,
\end{equation}
neglecting finite lifetimes or light-cone effects
\citep[see][for further discussion of these issues]{Croft04}.
Here $r_0\simeq \mathcal{O}(100\,h^{-1}{\rm Mpc})$ is the `attenuation length'
of the IGM, which is a rapidly decreasing function of redshift.

\citet{MeiWhi04} find, for a $\Lambda$CDM cosmology, that the (comoving)
attenuation length is approximately
\begin{equation}
  r_0 \simeq 2\times 10^4\ (1+z)^{-3.2}\ h^{-1}\,{\rm Mpc}
\end{equation}
for $2.75<z<5.5$, with diffuse gas and Lyman Limit Systems contributing
approximately equally.  This gives a comoving attenuation length of
$r_0\approx 200\,h^{-1}$Mpc at $z=3$, with larger values at lower $z$.
This result appears to agree with more recent measurements, for example
the recent measurement of \citet{ProWorMea09} at higher $z$ agree with
the results above where they overlap.
In \citet{SHWL09} a value of $r_0=100\,h^{-1}$Mpc was chosen, to emphasize
effects at the acoustic scale.  Here we use $r_0\approx 200\,h^{-1}$Mpc,
which is closer to the value inferred above, neglecting the dependence on
redshift.  Increasing $r_0$ suppresses the photo-ionization fluctuation
auto-correlation function on small scales while increasing it on the (large)
scales of interest (in agreement with analytic arguments; \citealt{Zuo92}).
This makes the impact of a fixed-rms $\Gamma$ fluctuation larger at the
acoustic scale, further decreasing the contrast of the acoustic peak.
For simplicity we have made the approximation that $r_0$ is the same for
all quasars.

To the particular realization of the QSO component we add a uniform piece (to
model the emission from faint AGN, galaxies and the IGM itself; \citealt{Mei09})
such that the rms fluctuation of the total is $10\%$ and divide our original
$\tau$ in every pixel by the ionization rate at that location.
The overall normalization is rescaled to match the mean flux, $\bar{F}$.

Ideally we would have used the positions of the dark matter halos for the
sources of our photo-ionizing flux.  However we do not have halo information
for the Roadrunner simulations.  To test for the possible effect of quasar
clustering on our $\Gamma(\vec{x})$ we looked at another set of simulations,
run using a high force-resolution TreePM code \citep{TreePM} in
$1.25\,h^{-1}$Gpc cubes, for which we did have stored halo catalogs.
Using the model of \citet{Cro09} we assigned quasar luminosities to each halo,
with the lowest luminosity quasar having $L/L_\star\simeq 2\%$.
The resulting luminosity function and clustering agree well with observations
at $z\sim 2-3$.
For both these `quasars', and a set with randomized positions but the same
luminosities, we computed $\Gamma(\vec{x})$ (as above) for $10^4$ randomly
placed skewers and hence $\xi_\Gamma$.  Comparison of $\xi_\Gamma$ for the
halo-produced $\Gamma(\vec{x})$ and the position-randomized $\Gamma(\vec{x})$
gives us a measure of the effect of source clustering.
There is significant variation, box-to-box, in $\xi_\Gamma$ but the mean at
$r\simeq 100\,h^{-1}$Mpc is increased by quasar clustering by $\sim 30(20)\%$
when $r_0=100(200)\,h^{-1}$Mpc.
These are small increases in $\xi_\Gamma$, but the fact that the halos trace
the density field induces correlations between $\Gamma$ and $\delta_m$ which
are missed by our random-position approximation and which we cannot adequately
probe.  It remains an open question whether matter-correlated source clustering
is important for $\Gamma$ fluctuations.
We also note that $r_0$ is becoming an appreciable fraction of the simulation
box length in our main simulations, raising questions of numerical convergence.
It would be nice to confirm and extend this analysis with larger simulations
including the halo catalogs, and we plan to return to this in a future
publication.

\subsection{Fluctuations in the IGM temperature}

In addition to the ionizing background fluctuations described above, the
Universe undergoes HeII reionization by bright QSOs around $z\sim 3$
\citep{Mei09}.
This event can heat regions of the IGM by as much as $25,000\,$K, with large
temperature fluctuations on $50\,$Mpc scales \citep{McQ09}.
The large QSO bubbles responsible for HeII reionization are essentially
uncorrelated with the overdensity at a given location and no value of $\delta$
is preferentially ionized at a given time during HeII reionization
\citep{McQ09}.  This leads to significant scatter in $T$ at fixed $\delta$
which is approximately ``random''.

In order to mock up such a situation in as simple a manner as possible, we
randomly throw $1,000$ centers within the box and assign each a radius of
influence equal to the mean inter-center separation and a temperature
drawn from a log-normal of mean $2\times 10^4\,$K and $\sigma_{\ln T}=1$.
The temperature at mean density at any point in the box is then the weighted
sum of these values, with a Gaussian weight depending on distance from each
center.
This gives fluctuations in the temperature at mean density which are
large, but highly coherent over large distances and smoothly varying.

\subsection{Effects on $\Delta\xi_F$}

\begin{figure}
\begin{center}
\resizebox{3.2in}{!}{\includegraphics{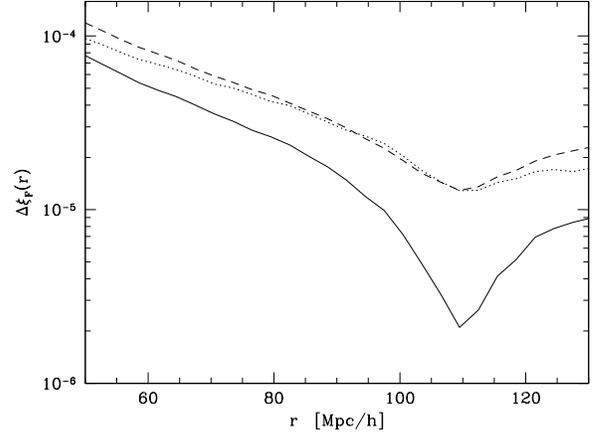}}
\end{center}
\caption{The compensated flux correlation function, $\Delta\xi_F$, vs.~scale
for our fiducial model (solid), a model with $\Gamma$-fluctuations as
described in the text (dotted) and a model with fluctuating $T_0$ (dashed).
The large ``extra'' power which dominates $\xi_F$ in the latter two models is
removed in $\Delta\xi_F$, and the acoustic signal remains visible though
reduced in contrast.}
\label{fig:jfluct}
\end{figure}

Both the fluctuations in the photo-ionization rate and modulation of the mean
temperature give rise to large changes in the correlation function, similar
to what was seen for an evolving $\bar{F}$ (\S\ref{sec:deltaXi}).
As there, much (but not all) of the `extra' signal is smooth and can be
removed by considering a compensated statistic.
Figure \ref{fig:jfluct} shows the effect of $\Gamma$ and $T_0$
fluctuations on $\Delta\xi_F$.

\subsection{Diagnostics}

Diagnostics of these non-gravitational contributions can be found in the
higher moments of the flux \citep{ZalSelHui01,FanWhi04}
and all indications are that the forest is dominated by gravitational
instability on large scales.  However, the errors on such measurements are
still large enough that the issue is not settled.

The existing measurements of inter-scale correlations were done on
individual spectra by computing the cross-power between a power of
the (band-pass) filtered flux field and the original flux field.
The auto- and cross-power spectra can be related to the power spectrum,
bispectrum and trispectrum of the flux.
Gravitational instability induces a positive correlation between large- and
small-scale density fluctuations, which leads to a negative correlation
coefficient, $C(k)$, between the filtered field and the original flux.

In the absence of non-gravitational contributions $C(k)\approx 0$ on scales
smaller than the cut-off in the band-pass filter and quickly tends to negative
values on larger scales.  The value of $C(k\to 0)$ depends primarily on the
width of the band-pass filter, being more negative the wider the filter.
Existing observations of a handful of high S/N, high resolution spectra are
consistent with the structure produced in gravity-only simulations using the
FGPA.

We find that in the presence of $\Gamma$-fluctuations the cross-correlation,
$C(k)$, is altered by a small amount at $k^{-1}\sim 100\,h^{-1}$Mpc for
band-passes focused on small scales.
Such an alteration would have been totally unmeasurable in the existing
analyses, and whether it is observable in the presence of unknown continuum
structure and other observational non-idealities remains in doubt.
The effects of large-scale coherent fluctuations in $T_0$ are also quite small
in this statistic and would not have been seen in existing analyses.

The basic idea of measuring the higher order correlations and looking
for gravitationally induced signatures remains valid however.  With
upcoming BOSS data as a motivation, we here present preliminary results
on the large-scale 3-point cross-correlation function along three
different lines-of-sight.
Such a statistic will be less susceptible to noise and continuum fluctuations
than the measurement discussed above.

\begin{figure}
\begin{center}
\resizebox{3.2in}{!}{\includegraphics{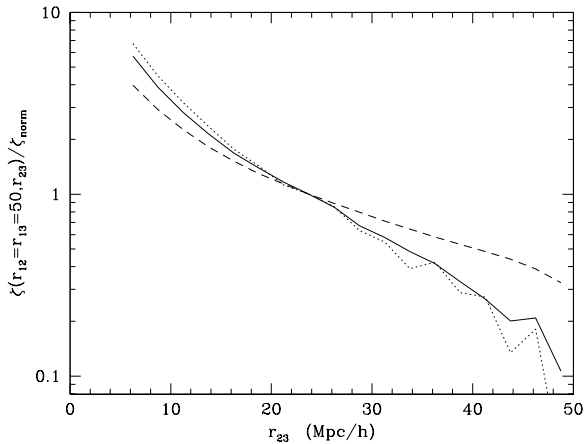}}
\end{center}
\caption{The 3-point flux correlation function in configuration space for
isosceles triangles with $r_{12}=r_{13}=50\,h^{-1}$Mpc vs.~$r_{23}$ for the
fiducial model (solid), a model with $\Gamma$-fluctuations as described in
the text (dotted) and a model with fluctuating $T_0$ (dashed).  An overall
amplitude has been removed by normalizing to $\zeta$ with $r_{23}=r_{12}/2$.
Note that the model with fluctuating $T_0$ shows a different scale dependence
than the other models.}
\label{fig:zeta}
\end{figure}

There are many possible configurations for which we could evaluate the flux
3-point function, and we could evaluate it in configuration or Fourier space.
We choose here to use an isosceles triangle in configuration space.
Particularly for $T_0$ fluctuations from He reionization, we might expect
configurations where two vertices are in one bubble and one is in another to
show a large non-gravitational signal.
Figure \ref{fig:zeta} shows the 3-point flux correlation function in
configuration space for isosceles triangles with $r_{12}=r_{13}=50\,h^{-1}$Mpc
vs.~$r_{23}$.
Our simulations show that there is a large variance in the 3-point function
{}from box to box, and smaller equations-of-state, $\gamma$, lead to a steeper
dependence of $\zeta$ on $r_{23}$.
The model with fluctuating $\Gamma$ cannot be distinguished from pure
gravity models with $1<\gamma<1.5$ based on the shape dependence shown here.
Similarly increasing the temperature uniformly to $3\times 10^4\,$K gives a
result close to the solid line in Figure \ref{fig:zeta}.
However, the model with $T_0$ fluctuations shows a much flatter dependence on
$r_{23}$ than the others, enabling a test of the non-gravitational nature of
the fluctuations.
We have chosen to plot $r_{12}=50\,h^{-1}$Mpc since on smaller scales the
3-point function is very similar for all of the models (as expected) while on
large scales the results become very noisy though the model differences appear
to be accentuated.
In principle, a survey like BOSS would be able to detect the difference
with high significance.
Whether these differences can be determined in the presence of other sources
of noise and modeling uncertainty, how the signals scale with the model
parameters and what the best higher-order statistic is, all need to be
investigated and we hope to return to this in future work.

\section{Discussion and Conclusions} \label{sec:conclude}

Baryon acoustic oscillations have become one of the premier methods
for determining cosmological distances and hence the expansion
history of the Universe.  The structure in the spectrum of distant
quasars, which is thought to trace the structure of the IGM at near
mean density, has been suggested as a relatively cheap method for
measuring BAO at high redshift.
Motivated by the upcoming BOSS experiment, and the availability of
the world's fastest supercomputer, ``Roadrunner'', we have performed
a set of ultra-large particle-mesh simulations in order to study the
BAO signature in the inter-galactic medium.

{}From the density and velocity fields we have used the ``fluctuating
Gunn-Peterson approximation'' (FGPA) to produce mock \lya spectra.
These simulations are the first to be able to cover the large volume
necessary to constrain the acoustic scale while simultaneously resolving
the Jeans scale of the gas, and the small-scale power and pixel
distribution in our mock spectra approximately match those seen in
observations at $z\simeq 2.5$.  We anticipate that these mock spectra
will be very useful in testing pipelines, calibrating analysis tools
and planning future projects so we have made them publicly
available\footnote{{\tt http://mwhite.berkeley.edu}}.

In the simulations the \lya flux follows the mass fluctuations on
large scales, with negligible scale-dependent bias \citep{SHWL09}.
This is in some sense to be expected.  We have assumed the FGPA in
producing our mock spectra and within this approximation the flux
is a highly non-trivial but deterministic function of the underlying
density field.

Since the small-scale statistics in our simulations approximately match
observations, we are able to make a reasonable estimate of the error in
the flux auto-correlation function.  We discuss the scaling of the error
with the number of quasar sightlines and the noise in each spectrum.
In the limit of low noise and many sightlines the error on the correlation
function is well approximated by the Gaussian expression.

We have presented preliminary investigations of an evolving mean flux,
fluctuations in the photo-ionization rate and HeII reionization which
generate ``extra'' power on the acoustic scale and reduce the contrast
of the acoustic peak.
Gravitational instability produces a well-defined pattern of higher order
correlations which is not obeyed by non-gravitational contributions such
as the above, allowing (in principle) a diagnostic of non-gravitational
physics in the forest.
As an example, we demonstrate that the 3-point cross-correlation function
in models with HeII reionization has a different scale dependence than the
3-point function in gravity-only simulations, regardless of the
equation-of-state assumed in the latter.

\vspace*{1cm}

\begin{acknowledgments}
  MW thanks Shirley Ho, Matt McQuinn, Avery Meiksin and Anze Slosar for
  conversations on the \lya\ forest.
  The simulations used in this work were run at LANL under the Roadrunner Open
  Science Program.
  Part of the analysis was done under the Los Alamos Institutional Computing
  Program.
  We thank the LANL Roadrunner support team for their outstanding effort in
  helping complete the simulation runs for this project.
  The analysis also made use of the computing resources of the
  National Energy Research Scientific Computing Center and the
  Laboratory Research Computing project at
  Lawrence Berkeley National Laboratory.
  MW is supported by the NASA and the DoE.
  Part of this research was supported by the DOE under contract W-7405-ENG-36.
  DD, PF, SH, KH, ZL, AP acknowledge support from the LDRD program at
  Los Alamos National Laboratory.
\end{acknowledgments}

\appendix

\section{Scaling with number density}

The distribution of quasars on the sky determines at what places we sample
the underlying density field.  As the number density of quasars increases
so does the number of samples with which we can constrain $P(k)$ or $\xi(r)$,
and hence the fidelity of the measurement, but eventually we are limited by
the available modes within the survey volume.
Here we describe how that limit is approached for a Gaussian field, which is
a reasonable approximation on large scales (see \S\ref{sec:cov}).

To begin we consider a large periodic volume of side length $L$ and volume
(or area in 2D) $V$.  Our Fourier transform convention has
\begin{equation}
  \delta(\mathbf{k}) = V^{-1}\int_V dV\ \delta(\mathbf{r})
  \exp\left[i\mathbf{k}\cdot\mathbf{r}\right]
  \qquad , \qquad
  \delta(\mathbf{x}) = \sum_{\mathbf{k}} \delta(\mathbf{k})
  \exp\left[-i\mathbf{k}\cdot\mathbf{r}\right]
\end{equation}
both dimensionless and
$P(k)=V\left\langle\left|\delta(\mathbf{k})\right|^2\right\rangle$.
To model the finite sampling we multiply the density field,
$\delta(\mathbf{x})$, by a window function, $w(\mathbf{x})$, which is
non-zero only where the field is observed, leading to
\begin{equation}
  P(\mathbf{k}) = \sum_{\mathbf{k}'} P(\mathbf{k'})
    \left|w_{\mathbf{k}-\mathbf{k}'}\right|^2
\end{equation}
We shall approximate $w(\mathbf{x})$ as a sum of (randomly distributed)
2D $\delta$-functions, or $w(\mathbf{k})$ as a sum of plane waves.
If we average in a shell $S_i$ with
$k_i-\Delta k/2<|\mathbf{k}|<k_i+\Delta k/2$
we obtain
\begin{equation}
  \widehat{P}_i \equiv \frac{1}{N_i} \sum_{\mathbf{k}\in S_i}
    \sum_{\mathbf{k}',\mathbf{k}''}
    \delta_{\mathbf{k}'}^\star\delta_{\mathbf{k}''}
    w^\star_{\mathbf{k}-\mathbf{k}'}w_{\mathbf{k}-\mathbf{k}''}
\end{equation}
with mean
\begin{equation}
  \left\langle \widehat{P}_i \right\rangle
  = \frac{1}{N_i} \sum_{\mathbf{k}\in S_i}
    \sum_{\mathbf{k}'} P(k') \left|w_{\mathbf{k}-\mathbf{k}'}\right|^2
\end{equation}
and covariance
\begin{equation}
  {\rm Cov}\left[\widehat{P}_i,\widehat{P}_j\right] = \frac{2}{N_i N_j}
     \sum_{\mathbf{k}_1 \in S_i}
     \sum_{\mathbf{k}_2 \in S_j}
     \sum_{\mathbf{k}',\mathbf{k}''} P(k') P(k'')
     \ w_{\mathbf{k}_2-\mathbf{k}'}
       w_{\mathbf{k}' -\mathbf{k}_1}
       w_{\mathbf{k}_1-\mathbf{k}''}
       w_{\mathbf{k}''-\mathbf{k}_2}  \quad .
\label{eqn:CovP}
\end{equation}
where $N_i$ is the number of modes in the shell and we have explicitly
assumed $P$ depends only on $|\mathbf{k}|$ (e.g.~we have neglected redshift
space distortions).

If $w(\mathbf{x})$ is a sum of $n_{\rm los}$ $\delta$-functions, whose
positions we average over, the product of $w(\mathbf{k})$'s in
Eq.~(\ref{eqn:CovP}) becomes a sum of Kronecker $\delta$'s.
There are 4 kinds of terms (depending on the number of points which coincide)
which scale as $1$, $1/n_{\rm los}$, $1/n_{\rm los}^2$ and $1/n_{\rm los}^3$
respectively.
In 2D in the continuum limit and for narrow bins the result becomes
\begin{equation}
  {\rm Var}[\widehat{P}_i] = \frac{2}{N_i}\ P(k_i)^2
  \left( 1 + \frac{\sigma^2}{\bar{n}P(k_i)} \right)^2
\end{equation}
where $\sigma^2$ is the density variance and $\bar{n}$ is the line-of-sight
number density.
The ratio of the terms going as $1$ and $1/\bar{n}$ determines at what
number density we approach the sample variance limit.

In 3D the situation is slightly more complex because the window function has
no structure in the line-of-sight direction, which gives a $\delta$-function
in $k_{\rm los}$, and the spherical average involved in $\widehat{P}_i$ gives
more structure to the expression.
In the continuum limit, the approach to the sample variance limit is
governed by the ratio\footnote{Including the next-order correction involving
$v_2$ is important for quantitative accuracy.}
\begin{equation}
  \frac{v_1}{\bar{n}v_0} \to \frac{\pi}{\bar{n}\, k_iP(k_i)}
  \ \int \frac{d^3\,\mathbf{k}}{(2\pi)^3} P(k) W(k)
\end{equation}
where the $v_j$ are defined in Eq.~(\ref{eqn:Vjdef}) and $W(k)$ is unity
for $|\mathbf{k}|<k_i$ and $k_i/|\mathbf{k}|$ for $|\mathbf{k}|>k_i$.
For $\Lambda$CDM cosmologies the integral is $\mathcal{O}(0.01)$ on the scales
of interest.  The additional variance depends then upon
$\bar{n}_{\rm eff} P(k_i)/P_0$ where
$\bar{n}_{\rm eff}=k\bar{n}_{2D}\sim \bar{n}/\lambda$ and we have included
the $P_0$ factor to indicate that the ratio is independent of the normalization
of $P(k)$.
We note that this expression is qualitatively similar to, but not the same
as, the expression given in \citet[][Eq.~13]{McD07}.

For small $k_i$ the integral above can be approximated by the 1D power spectrum
(an approximation that becomes worse as $k\to\infty$).  With that approximation
the ratio can be written as $\bar{n}_{\rm crit}/\bar{n}$, with the critical
(2D) number density being
\begin{equation}
  \bar{n}_{\rm crit} \equiv \frac{\Delta_{1D}^2(k_i)}{k_iP(k_i)/\pi}
  \qquad {\rm where} \qquad
  \Delta_{1D}^2(k_i) \equiv
  k_i\ \int_{k_i}^\infty \frac{d^3\,\mathbf{k}}{(2\pi)^3}
  \frac{P(k)}{k}
\end{equation}
which depends on the shape of the power spectrum but not its normalization.
As expected, cosmologies with more small-scale power require a higher number
density of skewers to reach the sample variance dominated limit.
For our model $\bar{n}_{\rm crit}\sim 0.01\,h^2\,{\rm Mpc}^{-2}$, in agreement
with the trends seen in Figure \ref{fig:error}.

\bibliography{ms}{}
\bibliographystyle{apj}

\end{document}